\documentclass[11pt,a4paper]{article}

\def\AmSTeX{\leavevmode\hbox{$\mathcal A\kern-.2em\lower.376ex%
        \hbox{$\mathcal M$}\kern-.2em\mathcal S$-\TeX}}
\usepackage{graphicx}
\usepackage{amsmath}        
\usepackage{amssymb}
\usepackage{mathrsfs}       
\usepackage{bm}             
\usepackage[amsmath,thmmarks,hyperref]{ntheorem}
\newif\ifpdf \pdftrue
\ifx\pdfoutput\undefined \pdffalse \fi \ifx\pdfoutput\relax
\pdffalse \fi
\ifx\texonly\undefined\let\texonly\relax\fi
\ifx\endtexonly\undefined\let\endtexonly\relax\fi \texonly
  \let\htmlonly\iffalse
  \let\endhtmlonly\fi
\endtexonly

\htmlonly
  
\endhtmlonly
\texonly
  \usepackage{makeidx}

  \usepackage{multicol}
  

\endtexonly
\title{}
\author{\thanks{}}
\date{}
\begin{document}

\title{Annihilation Rates of Heavy $1^{--}$ S-wave Quarkonia in Salpeter Method}

\author{Hui-feng Fu, Xiang-jun Chen, Guo-Li Wang\footnote{gl\_wang@hit.edu.cn}\\
{\it \small  Department of Physics, Harbin Institute of
Technology, Harbin, 150001, China} }

\maketitle

\baselineskip=20pt
\begin{abstract}
Annihilation rates of vector $1^{--}$ charmonium and bottomonium
$^3S_1$ states $V \rightarrow e^+e^-$ and $V\rightarrow 3\gamma$, $V
\rightarrow \gamma gg$ and $V \rightarrow 3g$ are estimated in the
relativistic Salpeter method. In our calculations, special attention
is paid to the relativistic correction, which is important and can
not be ignored for excited $2S$, $3S$ and higher excited states. We
obtain $\Gamma(J/\psi\rightarrow 3\gamma)=6.8\times 10^{-4}$ keV,
$\Gamma(\psi(2S)\rightarrow 3\gamma)=2.5\times 10^{-4}$ keV,
$\Gamma(\psi(3S)\rightarrow 3\gamma)=1.7\times 10^{-4}$ keV,
$\Gamma(\Upsilon(1S)\rightarrow 3\gamma)=1.5\times 10^{-5}$ keV,
$\Gamma(\Upsilon(2S)\rightarrow 3\gamma)=5.7\times 10^{-6}$ keV,
$\Gamma(\Upsilon(3S)\rightarrow 3\gamma)=3.5\times 10^{-6}$ keV and
$\Gamma(\Upsilon(4S)\rightarrow 3\gamma)=2.6\times 10^{-6}$ keV.
\end{abstract}

\section{Introduction}

Annihilation decay of $1^{--}$ S-wave heavy quarkonia has been
extensively studied
\cite{Appelquist,Rujula,Celmaster,Lepage,Scharre,Kramer,Field}.
The interest in this study comes from several sources. First, the
annihilation amplitudes are related to the behavior of wave
function, enabling an understanding of the formalism of
inter-quark interactions. Further more, it can be a sensitive test
of potential models \cite{Godfrey}. Finally, the ratio of the
decay widths, e.g. $\Gamma(V\rightarrow
e^+e^-)/\Gamma(V\rightarrow 3g)$, is sensitive to the running
coupling constant $\alpha_s(\mu)$, where $V$ is a
 heavy quarkonium vector state and $\mu$ is the scale ($\mu=m_c$ for
$J/\psi$ and $\mu=m_b$ for $\Upsilon$), and may provide very useful
information for $\alpha_s$ at the heavy quark mass scale
\cite{Kwong,CKT}.

In our previous Letters \cite{Wang1}, two-photon and two-gluon
annihilation rates of $J^{PC}=0^{-+}$, $0^{++}$ and $2^{++}$
$c\bar{c}$ and $b\bar{b}$ states are computed with the
relativistic Salpeter method. Good agreement of the predictions
with other theoretical calculations and the available experimental
data is found. In the calculations, we found the relativistic
corrections are large and not negligible, especially for high
excited states, such as, the $2S$ and $3S$ states, because there
are node structures in wave functions of $2S$ and $3S$ states,
these cause large relativistic corrections even for heavy
quarkonium like bottomonium. So in the theoretical studies
concerning the highly excited states, a relativistic model is
required.

The annihilation decays of the vector $1^{--}$ states are different
from the C even states. Basically there are two types of
annihilation decay modes, which are $V\rightarrow \gamma^*
\rightarrow l^+l^-$ and $V\rightarrow 3\gamma,\ \gamma gg,\ 3g$.
These decay widths have been studied in non-relativistic limit and
found to be proportional to the square of the wave function at the
origin $|\psi(0)|^2$ \cite{Appelquist,Ore}. However, the decay rates
of many processes are subject to substantial relativistic
corrections \cite{CKT,Keung,Chiang}. In this Letter, we will
continue to study the annihilation decays of $J^{PC}=1^{--}$
 $c\bar{c}$ and $b\bar{b}$ states with the relativistic Salpeter method.

There are two sources of relativistic corrections
\cite{Lepage,Wang1}, one is the correction in relativistic
kinematics which appears in the decay amplitudes through a
well-defined form of relativistic wave function (i.e., not merely
through the wave function at origin); the other relativistic
correction comes via the relativistic inter-quark dynamics, which
requires a relativistic formalism to describe the interactions
among quarks and relativistic formalism to consider the transition
amplitude. To consider the relativistic corrections, we choose the
Salpeter method \cite{Salpeter2}, which is an instantaneous
version of Bethe-Salpeter method \cite{Salpeter1}. For the
equal-mass quarkonium, the non-instantaneous correction is very
small \cite{Chang}. For the annihilation amplitude, we choose
Mandelstam formalism \cite{Mandelstam}, which is well suited for
the computation of relativistic transition amplitude with
Bethe-Salpeter wave functions as input.

In Section 2, we give theoretical details for the annihilation
amplitude in Mandelstam formalism and the corresponding wave
function with a well-defined relativistic form. The decay width of
$V\rightarrow \gamma^* \rightarrow e^+e^-$ and $V\rightarrow
3\gamma,\ \gamma gg,\ 3g$ are formulated in this section. We will
show the numerical results and give discussions in Section 3.
\section{Theoretical Details}
\subsection{The $V\rightarrow e^+e^-$ decay}

\begin{figure}[htbp]
\centering
\includegraphics[width = 0.6\textwidth]{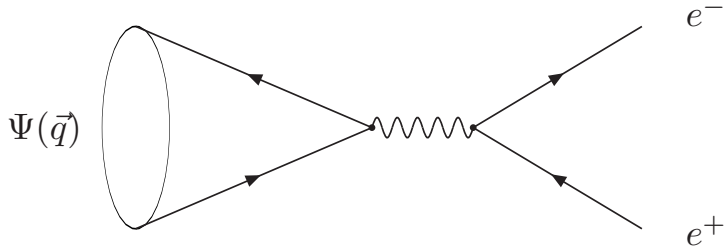}
\caption{Leptonic decay diagram of quarkonium.} \label{fig1}
\end{figure}

According to Mandelstam formalism \cite{Mandelstam}, the
transition amplitude of a quarkonium decaying into a electron and
a positron (see figure \ref{fig1}) can be written as

\begin{equation}
T_{e^+e^-}=i\sqrt{3}e^2e_q\int\frac{d^4q}{(2\pi)^4} \mathrm{tr}
[\chi(q)\gamma_\mu
]\frac{g^{\mu\nu}}{M^2}\bar{u}_{r_1}(\vec{k}_1)\gamma_\nu
v_{r_2}(\vec{k}_2),\label{eq1}
\end{equation}
where $e_q=\frac{2}{3}$ for charm quark and $e_q=-\frac{1}{3}$ for
bottom quark; $\vec{k}_1$ and $\vec{k}_2$ are the momenta of
electron and positron respectively; $M$ is the meson mass; $\chi(q)$
is the Bethe-Salpeter wave function with the total momentum $P$ and
relative momentum $q$, related by

$$p_1 = \alpha_1P + q,\  \alpha_1 \equiv \frac{m_1}{ m_1 + m_2} ,$$
$$ p_2 = \alpha_2P-q,\  \alpha_2\equiv\frac{ m_2}{ m_1 + m_2} ,$$
where $m_1=m_2$ is the constitute quark mass of charm or bottom.

After performing the integration over $q^0$, one reduce the
expression, with the notation of Salpeter wave function
$\Psi(\vec{q}) = i\int \frac{dq_0}{ 2\pi} \chi(q)$, to
\begin{equation}
T_{e^+e^-}=\sqrt{3}e^2e_q\int\frac{d\vec{q}}{(2\pi)^3} \mathrm{tr}
[\Psi(\vec{q})\gamma_\mu
]\frac{g^{\mu\nu}}{M^2}\bar{u}_{r_1}(\vec{k}_1)\gamma_\nu
v_{r_2}(\vec{k}_2).\label{eq2}
\end{equation}

We note that the form of the wave function is also important in
the calculation, since the corrections of the relativistic
kinetics come mainly through it. By analyzing the parity and
charge conjugation, the general form of relativistic wave function
of $1^{-}$ state ( $1^{--}$ for equal mass systems) can be written
as \cite{Wang2}
$$\Psi_{1^{-}}^{\lambda}(q_{\perp})=
q_{\perp}\cdot{\epsilon}^{\lambda}_{\perp}
\left[f_1(q_{\perp})+\frac{\not\!P}{M}f_2(q_{\perp})+
\frac{{\not\!q}_{\perp}}{M}f_3(q_{\perp})+\frac{{\not\!P}
{\not\!q}_{\perp}}{M^2} f_4(q_{\perp})\right]+
M{\not\!\epsilon}^{\lambda}_{\perp}f_5(q_{\perp})$$
\begin{equation}+
{\not\!\epsilon}^{\lambda}_{\perp}{\not\!P}f_6(q_{\perp})+
({\not\!q}_{\perp}{\not\!\epsilon}^{\lambda}_{\perp}-
q_{\perp}\cdot{\epsilon}^{\lambda}_{\perp})
f_7(q_{\perp})+\frac{1}{M}({\not\!P}{\not\!\epsilon}^{\lambda}_{\perp}
{\not\!q}_{\perp}-{\not\!P}q_{\perp}\cdot{\epsilon}^{\lambda}_{\perp})
f_8(q_{\perp}),\label{eq3}
\end{equation}
where $P$ and ${\epsilon}^{\lambda}_{\perp}$ are the momentum and
polarization vector of the vector meson; $q_\perp=(0,\vec{q})$. The
8 wave functions $f_i$ are not independent due to the equations
$\varphi^{+-}_{1^-}(q_\perp)=\varphi^{-+}_{1^-}(q_\perp)=0 $. For
quarkonium states we get the constraints on the components of the
wave functions \cite{Wang2}:
$$
f_1(q_\perp)=\frac{q_\perp^2 f_3(q_\perp)+M^2 f_5(q_\perp)}{M m_1},\
\  f_7(q_\perp)=0 ,
$$
$$
f_8(q_\perp)=-\frac{M f_6(q_\perp)}{ m_1},\ \ \ \  f_2(q_\perp)=0 .
$$
With these constraints, only four independent components $f_3, \
f_4,\ f_5$ and $f_6$ are left. Namely
$$\Psi_{1^{--}}^{\lambda}(q_{\perp})=
q_{\perp}\cdot{\epsilon}^{\lambda}_{\perp}(\frac{q_\perp^2}{M
m_1}+\frac{{\not\!q}_{\perp}}{M})f_3(q_\perp)+q_{\perp}\cdot{\epsilon}^{\lambda}_{\perp}\frac{\not\!P\not\!q_\perp}{M^2}f_4(q_\perp)
$$
\begin{equation}+
(M{\not\!\epsilon}^{\lambda}_{\perp}+q_{\perp}\cdot{\epsilon}^{\lambda}_{\perp}\frac{M}{m_1})f_5(q_\perp)
+[{\not\!\epsilon}^{\lambda}_{\perp}\not\!P+\frac{\not\!P(q_{\perp}\cdot{\epsilon}^{\lambda}_{\perp})}{m_1}
-\frac{\not\!P{\not\!\epsilon}^{\lambda}_{\perp}\not\!q_\perp}{m_1}]f_6(q_{\perp}).\label{eq4}
\end{equation}

These wave functions and the bound state mass $M$ can be obtained by
solving the full Salpeter equation with the constituent quark mass
as input. We will not show the details of how to solve the full
Salpeter equation, only give the final results. Interested readers
can find the detail technique in Refs.~\cite{Wang2,Wang3}.

Defining the decay constant $f_V$ by
\begin{eqnarray}
f_VM{\epsilon}^{\lambda}_\mu\equiv&\langle0|\bar{q_1}\gamma_\mu
q_{2} |V,\epsilon\rangle
&=\sqrt{3}\int\frac{d^3q}{(2\pi)^3}\mathrm{tr}[\varphi(\vec{q})\gamma_\mu],\label{eq6}
\end{eqnarray}
and with the Eq. (\ref{eq4}) we can easily obtain
\begin{eqnarray}
f_{V} = 4\sqrt{3} \int \frac{d \vec{q}}{(2\pi)^3} \left[f_{5}({\vec
q})-\frac{{\vec q}^2}{3M^2}f_3(\vec{q})\right].\label{eq7}
\end{eqnarray}

Summing over the polarizations of the final states and averaging
over that of the initial state, neglecting the electron mass, it is
easy to get the decay width
\begin{equation}
\Gamma_{e^+e^-}=\frac{4\pi}{3}\alpha^2 e_q^2 f_V^2 / M.\label{eq8}
\end{equation}

\subsection{$V\rightarrow 3\gamma$, $V \rightarrow \gamma gg$ and $V
\rightarrow 3g$ decays} 
\begin{figure}[htbp]
\centering
\includegraphics[width = 0.9\textwidth]{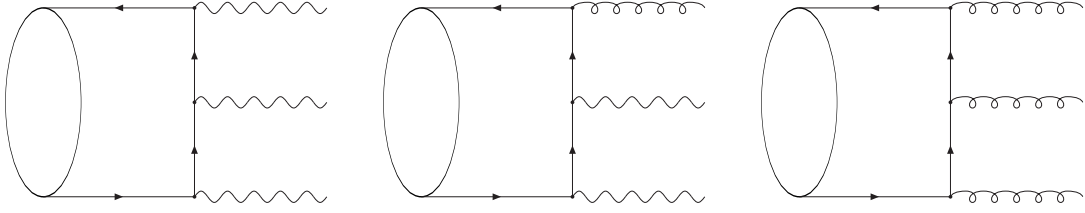}
\caption{$3\gamma,\ \gamma gg,\ 3g$ annihilation diagrams of
quarkonium. We do not show other diagrams with all the possible
permutations of photons and gluons.} \label{fig2}
\end{figure}

With the notation and definition used in
the previous subsection, the relativistic transition amplitude of
a quarkonium decaying into three photons (see figure \ref{fig2})
can be written as

$$
T_{3\gamma}=\sqrt{3}(iee_q)^{3}\int\frac{d^4q}{(2\pi)^4} \mathrm{tr}
\{\chi(q)[\not\!\epsilon_3
\frac{1}{\not\!k_3-\not\!p_2-m}\not\!\epsilon_2
\frac{1}{\not\!p_1-\not\!k_1-m}\not\!\epsilon_1
$$
\begin{eqnarray}
+\mathrm{all\ \ other \ \ permutations\ \ of \ \
1,2,3}]\},\label{eq9}
\end{eqnarray}
where $k_1,k_2,k_3$ and $\epsilon_1,\epsilon_2,\epsilon_3$ are the
momenta and polarization vectors of three photons respectively.

Since $p_{10} + p_{20} = M$, we assume $ p_{10} = p_{20} = M/2$ as
did in Ref. \cite{Wang1}. Having this assumption, we can perform
the integration over $q_0$ to reduce the expression to

$$
T_{3\gamma}=\sqrt{3}(ee_q)^{3}\int\frac{d^3\vec{q}}{(2\pi)^3}
\mathrm{tr} \{\Psi(\vec{q})[\not\!\epsilon_3
\frac{1}{\not\!k_3-\not\!\tilde{p}_2-m}\not\!\epsilon_2
\frac{1}{\not\!\tilde{p}_1-\not\!k_1-m}\not\!\epsilon_1
$$
\begin{equation}+\mathrm{all\
\ other \ \ permutations\ \ of \ \ 1,2,3}]\},\label{eq10}
\end{equation}
where $\tilde{p}_1=(\frac{M}{2},\vec{q}),
\tilde{p}_2=(\frac{M}{2},-\vec{q})$.

The decay width is given by
\begin{equation}
\Gamma_{3\gamma}=\frac{1}{3!}\frac{1}{8M
(2\pi)^3}\int_0^{\frac{M}{2}}dk_1^0
\int_{\frac{M}{2}-k_1^0}^{\frac{M}{2}}dk_2^0\
\frac{1}{3}\sum_{\mathrm{pol}}|T_{3\gamma}|^2\label{eq11}
\end{equation}

The width of $V \rightarrow \gamma gg$ and $V \rightarrow 3g$ are
related to the three photon decay width by
\begin{equation}
\Gamma_{\gamma gg}=\frac{2}{3}\frac{\alpha_s^2}{\alpha^2 e_q^4
}\Gamma_{3\gamma},\label{eq12}
\end{equation}
\begin{equation}
\Gamma_{3g}=\frac{5}{54}\frac{\alpha_s^3}{\alpha^3 e_q^6
}\Gamma_{3\gamma}.\label{eq13}
\end{equation}
For gluonic decay $V \rightarrow 3g$, the trace of color generators
gives $\mathrm{tr}[T_aT_bT_c]=\frac{1}{4}(d_{abc}+if_{abc})$, so the
expression of decay width contains two parts, one of which is
proportional to the square of symmetric constants and the other is
proportional to the square of antisymmetric constants. The existence
of antisymmetric term breaks the relation Eq.(\ref{eq13}). However,
our calculation shows that the antisymmetric term is sufficiently
small compared to the symmetric term, so we can ignore it safely. In
non-relativistic limit the antisymmetric term vanishes exactly.

\section{Numerical Results and Discussions}

To solve the full Salpeter equation, we choose a phenomenological
Cornell potential. There are some parameters in this potential
including the constituent quark mass and one loop running coupling
constant. The following best-fit values of input parameters were
obtained by fitting the mass spectra for heavy quarkonium $1^{--}$
states \cite{spectra}:
\begin{eqnarray}
~m_c=1.62 ~\mathrm{GeV},~ m_b=4.96 ~\mathrm{GeV}
\end{eqnarray}
For $c\bar{c}$ system, we set $\Lambda_{QCD}=0.27~\mathrm{GeV}$.
With this parameter set, we solve the full Salpeter equation and
obtain the mass spectra shown in Table \ref{tab1}. To give numerical
results , we need to fix the value of the renormalization scale
$\mu$ in $\alpha_s(\mu)$. In the case of charmonium, we choose the
charm quark mass $m_c$ as the energy scale and obtain the coupling
constant $\alpha_s(m_c) = 0.38$ \cite{Wang1}.

For $b\bar{b}$ system we set $\Lambda_{QCD}=0.20~\mathrm{GeV}$. With
this parameter set, the coupling constant at the scale of bottom
quark mass is $\alpha_s(m_b)=0.23$ \cite{Wang1}. The mass spectra
are also shown in Table \ref{tab1}.

With the obtained wave function, Eq.~(\ref{eq7}) and
Eq.~(\ref{eq8}), we calculate the decay width of $V\rightarrow
 e^+e^-$ for $c\bar{c}$ system. The results,
with other theoretical predictions and experimental data from
Particle Data Group, are shown in Table \ref{tab2}. Our results are
larger than experimental data and consistent with the Beyer's
\cite{Beyer} model version b results and Li's results~\cite{LC}. The
discrepancy between ours and experiment's may be due to the QCD
corrections. We only have the leading order QCD correction
$1-\frac{16}{3}\frac{\alpha_s}{\pi}$ \cite{Celmaster} in hand, while
the large factor $\frac{16}{3}$ implies that high order QCD
corrections can be still large and quite essential
\cite{Voloshin,Barbieri}, so we only show the results without QCD
corrections.

Decay widths of $\Upsilon(\textrm{nS})\rightarrow e^+e^-$ are shown
in Table \ref{tab3}. All the results, with or without QCD
corrections, are consistent with each other, with only small
discrepancies. Since the small value of $\alpha_s$ at the energy
scale of bottom quark, corresponds to much smaller QCD corrections
in bottomonium states than those in charmonium states, less
discrepancies exit among the results of $\Upsilon$(nS) decays than
of $\psi$(nS) decays.

The ratios of the high excited-state widths to the ground-state
width $\Gamma$(nS)$/\Gamma$(1S) are free from the QCD corrections
and sensitive to wave functions. We show the ratios of leptonic
decay widths in Table \ref{tab35}. Our theoretical values are
comparable to the PDG data, except those in $\psi(3S)$ and
$\psi(4S)$ states. It is can be seen from the table that the
ratios, so do the decay widths, fall very slowly with successive
radial excitations, which indicates that the relativistic
corrections are large for high excited states.

Decay widths $\Gamma_{3\gamma},\ \Gamma_{\gamma gg}~\rm and~
\Gamma_{3g}$ of charmonia and bottomonia are calculated with
Eqs.~(\ref{eq10}$\sim$\ref{eq13}). The results and other theoretical
estimates as well as experimental data are shown in Table \ref{tab4}
and Table \ref{tab5}. The decay widths quoted from Ref.~\cite{Kwong}
and Ref.~\cite{Voloshin} are estimated based on experimental data.
As in the $e^+e^-$ decays case, we only have the leading order QCD
correction, e.g. $1-12.6\frac{\alpha_s}{\pi}$ \cite{Celmaster}, in
hand, and the large factor $12.6$ implies that if we consider the
QCD corrections, we need include high order QCD corrections not only
the leading one. Besides, current leading order QCD factor is too
sensitive to the value of $\alpha_s$, which make other
contributions, such as relativistic corrections, unclear, so we only
show the results without QCD corrections. For the same reason in the
leptonic decays case, we show the ratios $\Gamma_{3
g}$(nS)/$\Gamma_{3 g}$(1S) in Table \ref{tab36}.

A typical non-relativistic calculation gives
$\mathcal{B}(J/\psi\rightarrow 3\gamma)\sim 3\times 10^{-5}$
\cite{Segovia,Voloshin}, while our relativistic result
$\mathcal{B}(J/\psi\rightarrow 3\gamma)\sim 0.73\times 10^{-5}$ (The
total width of $J/\psi$ is $93.2\pm2.1$ keV~\cite{Yao}) is much
smaller than the non-relativistic one, but within the experimental
error bar. This indicates that the relativistic corrections for
charmonia $3\gamma$ decays, so do the $\gamma gg,\ 3g$ decays, are
large. This conclusion is also obtained by other authors, see
Refs.~\cite{CKT,Keung,Chiang}. Besides, our calculations show that
the relativistic corrections for higher excited states are even
lager than those for the ground state and lower excited states.

One can see from Tables~\ref{tab4}$\sim$\ref{tab36}, that the decay
widths, which are sensitive to the wave functions of corresponding
states, fall very slowly from $1S$ to $5S$. We obtained the similar
results as the cases of $e^+e^-$ decays. This behavior is different
from the non-relativistic models, where the values fall quickly from
$1S$ to $5S$. It shows that the relativistic corrections are large
and important, especially for the higher excited states. It is
believed that the relativistic corrections are small for
bottomonium, however, we point that, this is true for ground state,
but not exactly true for the excited states, especially for high
excited states.

In calculating the decay widths $\Gamma_{3\gamma}$, we assume $
p_{10} = p_{20} = M/2$ . To show the effect of relaxing this
assumption, we take $ p_{10} =0.9\times M/2,\  p_{20} =1.1\times
M/2$ and estimate the relative deviations of decay widths
$(\Gamma-\Gamma_0)/\Gamma_0$, which are shown in
Table~\ref{tab37}. We interchange the values of $ p_{10}$ and $
p_{20} $, say, take $p_{10} =1.1\times M/2,\  p_{20} =0.9\times
M/2$, and find that the results are exactly the same as the
unchanged case as expected.

In summary, by solving the relativistic full Salpeter equation
with a well defined-form of wave function, we estimate
annihilation decay rates of heavy quarkonium  $1^{--}(^3S_1)$
states including $V\rightarrow  e^+e^-$, $V\rightarrow 3\gamma$,
$V\rightarrow \gamma gg$ and $V\rightarrow 3g$. We conclude that
the relativistic correction and QCD correction in these
annihilation decays play important roles, and high order QCD
corrections are expected.

\begin{table}[ph]
\caption{Mass spectra of the $c\bar{c}$ and $b\bar{b}$
$1^{--}(^3S_1)$ states ($^3D_1$ states are not presented) in unit
of MeV. The experimental data are taken from PDG \cite{Yao}.}
\begin{center}
{\begin{tabular}{|c|c|c|c|c|c|} \hline
$\psi(\mathrm{\mathbf{n}}S)$&Th($c\bar{c}$)&Ex($c\bar{c}$)&$\Upsilon(\mathrm{\mathbf{n}}S)$&Th($b\bar{b}$)&Ex($b\bar{b}$)
\\\hline  $J/ \psi$ &3096.9&3096.916&$\Upsilon$(1S)&9460.3&9460.30\\\hline$\psi$(2S)
&3688.4&3686.09&$\Upsilon$(2S)&10024&10023.26\\ \hline $\psi$(3S) &4056.0&4039&$\Upsilon$(3S)&10371&10355.2\\
\hline $\psi$(4S)&4327.7&4421&$\Upsilon$(4S)&10635&10579.4\\ \hline $\psi$(5S) &4543.3&&$\Upsilon$(5S)&10853&10865\\
\hline
\end{tabular} \label{tab1}}
\end{center}
\end{table}

\begin{table}[ph]
\caption{Decay width $\Gamma(\psi(\rm{n}S)\rightarrow e^+e^-)$ in
unit of keV. The results marked by $\dag$ do not cover the
contributions of QCD corrections. }
\begin{center}
{\begin{tabular}{|c|c|c|c|c|c|} \hline
$\psi(\mathrm{\mathbf{n}}S)$&$J/\psi$&$\psi$(2S)&$\psi$(3S)&$\psi$(4S)&$\psi$(5S)
\\\hline  Ours$^\dag$ &10.9&5.22&3.49&2.61&2.07\\ \hline Beyer Ver.a$^\dag$~\cite{Beyer} &5.33&2.31&1.59&1.14&\\
\hline Beyer Ver.b$^\dag$~\cite{Beyer}&11.2&4.06&2.74&2.06&\\
\hline EQ$^\dag$~\cite{EQ} &8.00&3.67&&&\\ \hline
VPBK$^\dag$~\cite{VPBK}
&5.469&2.140&0.796&0.288&\\ \hline GVGV~\cite{GVGV} &2.94&1.22&0.76&0.43&0.27\\
\hline IS~\cite{IS} &$6.72\pm 0.49$&$2.66\pm0.19$&&&\\
\hline SYEF~\cite{Segovia}
&3.93&1.78&1.11&&\\ \hline LC$^\dag$~\cite{LC} &11.8&4.29&2.53&1.73&1.25\\
\hline
PDG~\cite{Yao} &$5.55\pm0.14\pm0.02$&$2.38\pm0.04$&$0.86\pm0.07$&$0.58\pm0.07$&\\
\hline
\end{tabular} \label{tab2}}
\end{center}
\end{table}

\begin{table}[ph]
\caption{Decay width $\Gamma(\Upsilon(\rm{n}S)\rightarrow e^+e^-)$
in unit of keV. The results marked by $\dag$ do not cover the
contributions of QCD corrections. }
\begin{center}
{\begin{tabular}{|c|c|c|c|c|c|} \hline
$\Upsilon(\rm{n}S)$&$\Upsilon(\rm{1}S)$&$\Upsilon(\rm{2}S)$&$\Upsilon(\rm{3}S)$&$\Upsilon(\rm{4}S)$&$\Upsilon(\rm{5}S)$
\\\hline  Ours$^\dag$ &1.47&0.736&0.530&0.425&0.359\\ \hline Beyer Ver.a$^\dag$~\cite{Beyer}&1.24&0.51&0.35&0.28&\\
\hline Beyer Ver.b$^\dag$~\cite{Beyer}&1.41&0.56&0.36&0.30&\\ \hline EQ$^\dag$~\cite{EQ}&1.71&0.76&0.55&&\\
\hline VPBK$^\dag$\cite{VPBK}&1.320&0.628&0.263&0.104&0.0404\\ \hline GVGV~\cite{GVGV}&0.98&0.41&0.27&0.20&0.16\\
\hline IS~\cite{IS}&$1.45\pm0.07$&$0.52\pm0.02$&$0.35\pm0.02$&&\\ \hline Gonz$\rm{\acute{a}}$lez~\cite{Gonzalez}&1.7&0.61&0.39&0.27&0.21\\
\hline PDG~\cite{Yao} &$1.340\pm0.018$&$0.612\pm0.011$&$0.443\pm0.008$&$0.272\pm0.029$&$0.31\pm0.07$\\
\hline
\end{tabular} \label{tab3}}
\end{center}
\end{table}

\begin{table}[ph]
\caption{The ratios of the high excited-state widths to the
ground-state width $\Gamma$(nS)$/\Gamma$(1S) for decay
$\Gamma(\psi(\rm{n}S)\rightarrow e^+e^-)$ and
$\Gamma(\Upsilon(\rm{n}S)\rightarrow e^+e^-)$.}
\begin{center}
{\begin{tabular}{|c|c|c|c|c|} \hline
$\Gamma(\psi(\rm{n}S))/\Gamma(\psi(\rm{1S}))$&$\Gamma(\rm{2S})/\Gamma(\rm{1S})$&$\Gamma(\rm{3S})/\Gamma(\rm{1S})$
&$\Gamma(\rm{4S})/\Gamma(\rm{1S})$&$\Gamma(\rm{5S})/\Gamma(\rm{1S})$
\\\hline  Ours &0.48&0.32&0.24&0.19\\
\hline PDG~\cite{Yao} &0.43&0.15&0.10&\\
\hline\hline
$\Gamma(\Upsilon(\rm{n}S))/\Gamma(\Upsilon(\rm{1}S))$&$\Gamma(\rm{2S})/\Gamma(\rm{1S})$&$\Gamma(\rm{3S})/\Gamma(\rm{1S})$
&$\Gamma(\rm{4S})/\Gamma(\rm{1S})$&$\Gamma(\rm{5S})/\Gamma(\rm{1S})$
\\\hline  Ours &0.50&0.36&0.29&0.24\\
\hline PDG~\cite{Yao} &0.46&0.33&0.20&0.23\\
\hline
\end{tabular} \label{tab35}}
\end{center}
\end{table}

\begin{table}[ph]
\caption{Decay widths of $\psi(nS)\rightarrow 3\gamma,\ \gamma gg,
\ 3g$ in unit of keV. The data marked by $*$ is quoted from
Ref.~\cite{Adams}. The results marked by $\dag$ do not cover the
contributions of QCD corrections. }
\begin{center}
{\begin{tabular}{|c|c|c|c|c|c|c|} \hline Decay& Ours$^\dag$
&GI$^\dag$~\cite{Godfrey}&ML~\cite{ML}&PCP$^\dag$~\cite{PCP}&SYEF~\cite{Segovia}&Voloshin~\cite{Voloshin}
\\\hline $\Gamma (J/\psi\rightarrow
3g)$&101&176&$80\pm40$&63.72&&$61.5\pm3.1$\\\hline $\Gamma
(\psi(\mathrm{2S})\rightarrow
3g)$&36.6&78.4&&20.49&&$45.3\pm9.3$\\\hline $\Gamma
(\psi(\mathrm{3S})\rightarrow 3g)$&24.7&&&11.92&&\\\hline $\Gamma
(\psi(\mathrm{4S})\rightarrow 3g)$&19.8&&&8.08&&\\\hline $\Gamma
(\psi(\mathrm{5S})\rightarrow 3g)$&16.7&&&&&\\\hline \hline $\Gamma
(J/\psi\rightarrow \gamma
gg)$&6.18&&$7.5\pm3$&&&$7.46\pm2.80$\\\hline $\Gamma
(\psi(\mathrm{2S})\rightarrow \gamma gg)$&2.25&&&&&3.04\\\hline
$\Gamma (\psi(\mathrm{3S})\rightarrow \gamma gg)$&1.52&&&&&\\\hline
$\Gamma (\psi(\mathrm{4S})\rightarrow \gamma gg)$&1.22&&&&&\\\hline
$\Gamma (\psi(\mathrm{5S})\rightarrow \gamma gg)$&1.03&&&&&\\\hline
\hline $\Gamma (J/\psi\rightarrow 3\gamma)$&$0.68\times
10^{-3}$&&&&$0.56\times 10^{-3}$&$^{*}(1.12\pm0.47)\times
10^{-3}$\\\hline $\Gamma (\psi(\mathrm{2S})\rightarrow
3\gamma)$&$0.25\times 10^{-3}$&&&&&\\\hline $\Gamma
(\psi(\mathrm{3S})\rightarrow 3\gamma)$&$0.17\times
10^{-3}$&&&&&\\\hline $\Gamma (\psi(\mathrm{4S})\rightarrow
3\gamma)$&$0.13\times 10^{-3}$&&&&&\\\hline $\Gamma
(\psi(\mathrm{5S})\rightarrow 3\gamma)$&$0.11\times 10^{-3}$&&&&&
\\\hline
\end{tabular} \label{tab4}}
\end{center}
\end{table}

\begin{table}[ph]
\caption{Decay widths of $\Upsilon(nS)\rightarrow 3\gamma,\ \gamma
gg, \ 3g$ in unit of keV. The results marked by $\dag$ do not
cover the contributions of QCD corrections. }
\begin{center}
{\begin{tabular}{|c|c|c|c|c|} \hline Decay& Ours$^\dag$
&GI$^\dag$~\cite{Godfrey}&ML~\cite{ML}&KMRR~\cite{Kwong}
\\\hline $\Gamma (\Upsilon(\mathrm{1S})\rightarrow
3g)$&32.5&44.1&$28\pm6$&$42.9\pm1.2$\\\hline $\Gamma
(\Upsilon(\mathrm{2S})\rightarrow 3g)$&12.0&22.5&&\\\hline $\Gamma
(\Upsilon(\mathrm{3S})\rightarrow 3g)$&7.47&16.9&&\\\hline $\Gamma
(\Upsilon(\mathrm{4S})\rightarrow 3g)$&5.52&12.1&&\\\hline $\Gamma
(\Upsilon(\mathrm{5S})\rightarrow 3g)$&4.41&&&\\\hline \hline
$\Gamma (\Upsilon(\mathrm{1S})\rightarrow \gamma
gg)$&0.826&&$0.9\pm0.2$&$1.20$\\\hline $\Gamma
(\Upsilon(\mathrm{2S})\rightarrow \gamma gg)$&0.304&&&\\\hline
$\Gamma (\Upsilon(\mathrm{3S})\rightarrow \gamma
gg)$&0.190&&&\\\hline $\Gamma (\Upsilon(\mathrm{4S})\rightarrow
\gamma gg)$&0.140&&&\\\hline $\Gamma
(\Upsilon(\mathrm{5S})\rightarrow \gamma gg)$&0.112&&&\\\hline
\hline $\Gamma (\Upsilon(\mathrm{1S})\rightarrow
3\gamma)$&$0.15\times 10^{-4}$&&&\\\hline $\Gamma
(\Upsilon(\mathrm{2S})\rightarrow 3\gamma)$&$0.57\times
10^{-5}$&&&\\\hline $\Gamma (\Upsilon(\mathrm{3S})\rightarrow
3\gamma)$&$0.35\times 10^{-5}$&&&\\\hline $\Gamma
(\Upsilon(\mathrm{4S})\rightarrow 3\gamma)$&$0.26\times
10^{-5}$&&&\\\hline $\Gamma (\Upsilon(\mathrm{5S})\rightarrow
3\gamma)$&$0.21\times 10^{-5}$&&&
\\\hline
\end{tabular} \label{tab5}}
\end{center}
\end{table}

\begin{table}[ph]
\caption{The ratios of the high excited-state widths to the
ground-state width $\Gamma$(nS)$/\Gamma$(1S) for decay
$\Gamma(\psi(\rm{n}S)\rightarrow 3g)$ and
$\Gamma(\Upsilon(\rm{n}S)\rightarrow 3g)$.}
\begin{center}
{\begin{tabular}{|c|c|c|c|c|} \hline
$\Gamma(\psi(\rm{n}S))/\Gamma(\psi(\rm{1S}))$&$\Gamma(\rm{2S})/\Gamma(\rm{1S})$&$\Gamma(\rm{3S})/\Gamma(\rm{1S})$
&$\Gamma(\rm{4S})/\Gamma(\rm{1S})$&$\Gamma(\rm{5S})/\Gamma(\rm{1S})$
\\\hline  Ours &0.36&0.24&0.20&0.17\\
\hline GI~\cite{Godfrey} &0.45&&&\\
\hline PCP~\cite{PCP} &0.32&0.19&0.13&\\
\hline Voloshin~\cite{Voloshin} &0.74&&&\\
\hline\hline
$\Gamma(\Upsilon(\rm{n}S))/\Gamma(\Upsilon(\rm{1}S))$&$\Gamma(\rm{2S})/\Gamma(\rm{1S})$&$\Gamma(\rm{3S})/\Gamma(\rm{1S})$
&$\Gamma(\rm{4S})/\Gamma(\rm{1S})$&$\Gamma(\rm{5S})/\Gamma(\rm{1S})$
\\\hline  Ours &0.37&0.23&0.17&0.14\\
\hline GI~\cite{Godfrey} &0.51&0.38&0.27&\\
\hline
\end{tabular} \label{tab36}}
\end{center}
\end{table}

\begin{table}[ph]
\caption{ Relative deviations of decay width $\Gamma_{3\gamma}$ with
assumptions  $ p_{10} =0.9\times M/2,\ p_{20} =1.1\times M/2$ from
that with $ p_{10} = p_{20} = M/2$. }
\begin{center}
{\begin{tabular}{|c|c|c|c|c|c|} \hline nS&1S&2S &3S&4S&5S
\\\hline  $c\bar{c}$&$-2.0\%$&$-2.9\%$&$-4.7\%$&$-7.0\%$&$-8.2\%$\\
\hline $b\bar{b}$&$-1.8\%$&$-2.0\%$&$-2.2\%$&$-2.4\%$&$-2.6\%$\\
\hline
\end{tabular} \label{tab37}}
\end{center}
\end{table}

\section*{Acknowledgments}

This work was supported in part by the National Natural Science
Foundation of China (NSFC) under Grant No. 10875032 and supported
in part by Projects of International Cooperation and Exchanges
NSFC under Grant No. 10911140267, and supported in part by the
Foundation of Harbin Institute of Technology (Weihai) No. IMJQ
10000076.

\end{document}